\documentclass{revtex4}  
\usepackage[latin1]{inputenc}
\usepackage{amsmath}
\usepackage{amsfonts}
\usepackage{amssymb}
\usepackage{graphicx}

\begin{document}

\title{3D-PSTD for modelling second harmonic generation in periodically poled lithium niobate ridge-type waveguides}
\author{Fabrice Devaux and Mathieu Chauvet}
\address{Institut FEMTO-ST, D\'epartement d'Optique P. M. Duffieux, UMR 6174 CNRS \\ Universit\'e Bourgogne Franche-Comt\'e, 15b Avenue des Montboucons, 25030 Besan\c{c}on - France}


\begin{abstract}We report an application of the tri-dimensional pseudo-spectral time domain algorithm, that solves with accuracy the nonlinear Maxwell's equations, to predict second harmonic generation in lithium niobate ridge-type waveguides with high index contrast. Characteristics of the nonlinear process such as conversion efficiency as well as impact of the multimode character of the waveguide are investigated as a function of the waveguide geometry in uniformly and periodically poled medium.    
\end{abstract}

\maketitle
\section{Introduction} 
Optical nonlinear devices based on ferroelectric materials such as Lithium Niobate ($LiNbO_3$) or Lithium Tantalate ($LiTaO_3$) have a great technological potential. These cristals indeed allow the generation of any wavelength from the mid IR to the UV using quasi phase-matched nonlinear processes \cite{Mizuuchi_2003,Tovstonog_2007}. This technology is usefull to realise optical souces or can also play a key role inside wavelength multiplexing systems for optical telecommunications. Highly nonlinear optical fibers are competing alternative thanks to their ability to convert wavelengths through four wave mixing process via the Kerr effect. However, weak third order nonlinear coefficients give weak conversion efficiency even over several hundred meters of propagation. To the contrary, efficient conversion efficiency are possible over short distance (few centimeters) in periodically poled Lithium Niobate (PPLN) due to a large second order nonlinear coefficient $\chi^{(2)}$ allowing fast parametric conversions. 

Particularly, ridge waveguides with high index contrast have a key role to play in the improvement of performances \cite{Kurimura_2006}. In comparison with standard waveguide fabrication techniques, such as proton exchange \cite{Chou_1999} or titanium in-diffusion \cite{Lee_2004}, stronger confinement can be reached with ridge waveguides along with a good overlap between fundamental modes of the nonlinear process thanks to a high index contrast. In addition, long term stability of the devices is improved especially if the nonlinear material is doped with proper elements such as magnesium to limit the photorefractive effect. At last, fabrication techniques based on micromachinig or etching keep the intrinsic properties of the material. In particular, the strong nonlinear coefficients and low optical absorption coefficient remain unchanged. However, high index contrasts implies very small waveguide cross sections (sub-micron square) in order to form singlemode waveguides. Such a strong confinment is beneficial to envision high nonlinear conversion efficiency but it implies tight tolerances on the fabrication process of the periodic poling for quasi-phasematching and waveguide geometrical uniformity. Moreover, efficient light coupling in these tiny structures is very challenging. As a consequence, an alternative solution is to realise waveguides with larger cross sections (from 5$\mu m^2$ to 100$\mu m^2$) which releases constrains both on the fabrication process and on light coupling. In that case, influence of the multimode character of these waveguides on the nonlinear process has to be studied.

To optimize the design and performances of these waveguides, numerical modeling of waves propagation in such devices is essential. Many numerical methods are available for wave propagation modeling. The Finite-Difference Time-Domain method\cite{Taflovebook} (FDTD), the Split-Step Method\cite{Agrawalbook} (SSM) and the Finite Element Method\cite{Koshibabook} (FEM) belongs to the time-domain methods while the Beam Propagation Method \cite{kawanobook} (BPM), the Transfer Matrix Method (TMM) and the Eigen Mode Expansion method (EME) work in the frequency-domain.
Among these numerical techniques, the FDTD (implemented in a large number of both free and commercially available sofware) is the most general and rigorous time-domain method. It provides solutions for a large number of guided optics configurations such as photonic crystal waveguides, surface plasmon waveguides, devices with high-index contrast waveguides, ring and disk resonators, negative index material structures, dispersive, and nonlinear materials. However, when nonlinear phenomena in long guiding structures are studied, the FDTD becomes prohibitively computationally intensive and therefore impractical.

In order to overcome this later limitation we developed a numerical model, based on a Pseudo-Spectral Time Domain algorithm\cite{liu_pstd_1997} (PSTD). The nonlinear Maxwell's equations is solved in three spatial dimensions to fully characterise the electromagnetic fields of the wave injected in the ridge waveguide and the generated harmonic wave. This paper is organised as follows. In section \ref{theorie}, the principle of the PSTD algorithm is described. Then, the  numerical results of second harmonic generation (SHG) in uniformly poled and in periodically poled ridge waveguides are given in section \ref{resultats}. These results are compared to basic analytical calculations in order to validate the numerical model. Finally, modal analysis of the guided waves is performed in order to study the influence of the multimode property of the ridge waveguides on the SHG efficiency.

\section{Numerical model}\label{theorie}
\begin{figure}[ht]
   \begin{center}   
   \includegraphics[width=8cm]{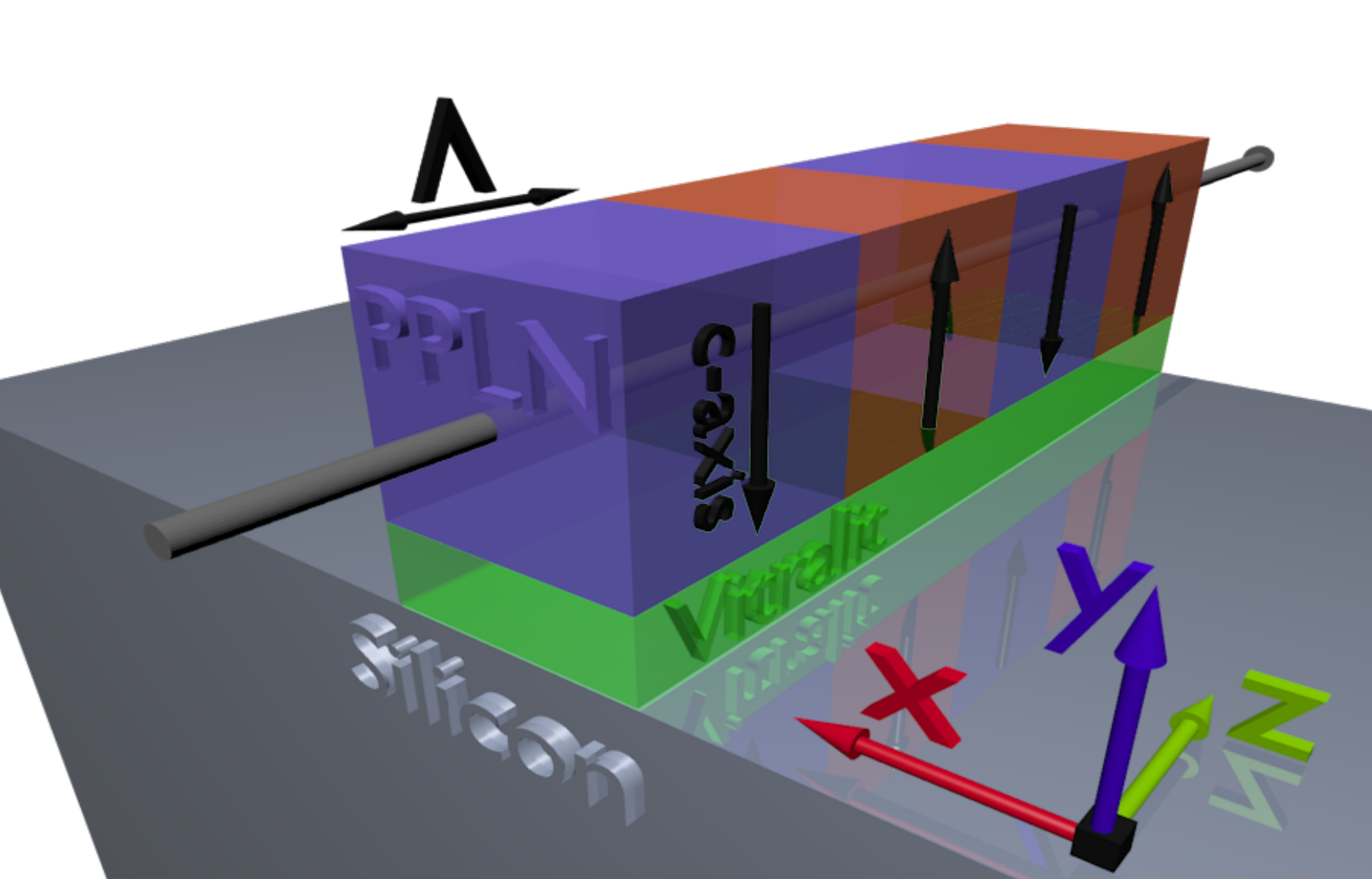}
   \end{center}
   \caption{ \label{schemaridgeSHG}
Schematic view of the modeled periodically poled lithium niobate ridge-type waveguide. The square section ridge is bonded to a silicon substrate with a buffer layer. $2\varLambda$ is the period of the poled domains and $z$ is the propagation axis of light.}
  \end{figure}
In PSTD algorithm, Maxwell's curl equations are calculated with discrete Fourier transforms in order to solve the spatial derivatives on an unstaggered, collocated grid \cite{huo_liu_review_2004}.This spatial differential process converges with infinite order of accuracy for grid-sampling densities of two or more points per wavelength \cite{liu_pstd_1997}, provided that the medium optical properties are sampled in accordance with the Nyquist theorem. Limitations of Fast Fourier Transform (FFT), due to the periodic boundary conditions, are avoided by using absorbing boundary conditions formulated for Perfect Matched Layer (PML) in nonconductive media \cite{liu_pstd_1997,berenger_perfectly_1994,chew_3d_1994}.
As a consequence, this numerical method can be used to study various problems on larger scales, more efficiently and with a better accuracy than Finite-Difference Time-Domain (FDTD) methods \cite{liu_pstd_1997,tang_2.5d_2003,TsengOE2007,TsengOE2009,devauxOpEx2013,CourjalOpEx2015}. In particular, because accurate modelling of nonlinear optical processes with the FDTD method requires extremely fine sampling to minimize numerical dispersion errors, PSTD schemes offer significant improvements in computational efficiency and accuracy \cite{lee_pseudospectral_2004,devauxPhotEurope2014}. In the present work, 3D-PSTD algorithm models the SHG in periodically poled and uniformly poled Lithium Niobate (LN) ridge-type waveguides. The modeled device is represented schematically in Fig. \ref{schemaridgeSHG}.  

In our numerical problem, we consider the propagation along the $z$ axis of two transverse, electromagnetic waves inside the ridge-type waveguide. More specifically we consider the fundamental wave and its second harmonic (SH) wave in CW regime. The $(x,y,z)$ axis correspond respectively to the $X,Z,Y$ axis of the LN crystal ($Z$ axis corresponds to the crystallographic c-axis of the LN crystal). The total volume is sampled with $n_x\times n_y\times n_z$ points giving a grid-sampling density greater than two points per the shortest considered wavelength (inside the LN crystal). The square section waveguide is bonded to a silicon substrate with a 3 $\mu m$ thick transparent buffer layer. $\Lambda$ is the length of the poled domains (Fig. \ref{schemaridgeSHG}) which gives a periodic poling of period $2\Lambda$.     
 
Use of FFT in Maxwell's curl equations yields time-stepping relations of the form given in reference \cite{devauxOpEx2013}. For example, the $x$ component of the electric displacement $\bold{D_f}$ and of the magnetic field $\bold{B_f}$ at the fundamental wavelength is expressed as:
\begin{eqnarray}\label{eq1} \left\{\begin{array}{cc}
D_{x,f}|^{n+1}_{jkl} = &
\frac{\left(1-\frac{\Delta
t}{2}\gamma^y_{jkl}\right)D_{xy,f}|^{n+\frac{1}{2}}_{jkl}+\frac{\Delta t}{\mu_0}F^{-1}_y\left(-2i\pi
\nu_yF_y\left(B_{z,f}|^{n}_{jkl}\right)\right)}{1+\frac{\Delta
t}{2}\gamma^y_{jkl}}+\frac{\left(1-\frac{\Delta
t}{2}\gamma^z_{jkl}\right)D_{xz,f}|^{n+\frac{1}{2}}_{jkl}-\frac{\Delta t}{\mu_0}F^{-1}_z\left(-2i\pi
\nu_zF_z\left(B_{y,f}|^{n}_{jkl}\right)\right)}{1+\frac{\Delta t}{2}\gamma^z_{jkl}}\\ \\
B_{x,f}|^{n+1}_{jkl} = & \frac{\left(1-\frac{\Delta
t}{2}\gamma^y_{jkl}\right)B_{xy,f}|^{n+\frac{1}{2}}_{jkl}-\Delta
tF^{-1}_y\left(-2i\pi \nu_yF_y\left(E_{z,f}|^{n}_{jkl}\right)\right)}{1+\frac{\Delta
t}{2}\gamma^y_{jkl}}+\frac{\left(1-\frac{\Delta
t}{2}\gamma^z_{jkl}\right)B_{xz,f}|^{n+\frac{1}{2}}_{jkl}+\Delta tF^{-1}_z\left(-2i\pi
\nu_zF_z\left(E_{y,f}|^{n}_{jkl}\right)\right)}{1+\frac{\Delta t}{2}\gamma^z_{jkl}}\\
\end{array}\right.\end{eqnarray}

where $D_{xy,f}$ is incremented by the partial derivative of $B_{z,f}$ with respect to $y$ and $D_{xz,f}$ is incremented by the partial derivative of $B_{y,f}$ with respect to $z$ \cite{devauxOpEx2013}. $(j,k,l)$ are the indices of the cartesian coordinates at the considered spatial point $(j\Delta x,k\Delta y,l\Delta z)$ and $n$ is the index of the temporal step $n\Delta t$. The
maximum time step to fullfill the stability criteria of the 3D-PSTD algorihm is $\Delta t=\frac{2\Delta x}{\pi c \sqrt{3}}$ \cite{huo_liu_review_2004}, where $c$ is the speed of light in vacuum.  $\gamma^v_{jkl}$ is the boundary absorbing layer function along the $v$ ($v\equiv x,y,z$) dimension which is obtained by setting the term $\gamma^v_{jkl}$ to zero inside the region of interest and to a pure imaginary part increasing linearly along the $v$ dimension inside the absorbing domains \cite{liu_pstd_1997} with widths of $\frac{n_v}{8}$ sampling periods at the boundaries of the sampled volume.  
$F_v$ and $F^{-1}_v$ correspond to the FFT and the inverse FFT along the dimension $v$.
Other components of $\bold{D_f}$ and $\bold{B_f}$ are obtained by circular permutation of the $x,y,z$ indices in Eq. \ref{eq1}. Similar equations give the fields components of the SH wave.

The source terms that generate the fundamental wave is designed such as the transverse electromagnetic
wave emitted by the source propagates along the $z$ direction. They are added at the intermediate time-step $n+\frac{1}{2}$, to the terms $D_{xz,f}|^{n}_{jkl}$ and $D_{yz,f}|^{n}_{jkl}$ using the relations :
\begin{eqnarray}\label{eq2} \left\{\begin{array}{cc}
D_{xz,f}|^{n+\frac{1}{2}}_{jkl} = &
D_{xz,f}|^{n}_{jkl}+S_{x,f}|^{n}_{jkl}\\ D_{yz,f}|^{n+\frac{1}{2}}_{jkl} = & D_{yz,f}|^{n}_{jkl}+S_{y,f}|^{n}_{jkl}\\
\end{array}\right. \end{eqnarray}
$S_{x,f}|^{n}_{jkl}$ and $S_{y,f}|^{n}_{jkl}$ are the amplitude of the transverse components of the CW source term at the time step $n\Delta t$ and at the spatial sampling point
$(j\Delta x,k\Delta y,l\Delta z)$. These components are given by:
\begin{eqnarray}\label{eq3} \left\{\begin{array}{cc}
S_{x,f}|^{n}_{jkl} = & S_{0,f}|_{jkl}\cos\psi_f e^{-i(\omega_f n\Delta t +\varphi_{x,f})}\\ S_{y,f}|^{n}_{jkl} = & S_{0,f}|_{jkl}\sin\psi_f e^{-i(\omega_f n\Delta t
+\varphi_{y,f})}\\
\end{array}\right.
\end{eqnarray}
$\psi_f$, $\varphi_{x,f}$ and $\varphi_{y,f}$ define the polarization state of the
electromagnetic wave emitted by the source. $S_{0,f}|_{jkl}$ gives a gaussian beam in the $(x,y)$ transverse plane and with the optimized three-cells normalized pattern
$[\frac{1}{4},\frac{1}{2},\frac{1}{4}]$ along the $z$ axis in order to suppress the aliasing errors \cite{lin_optimal_2010}.
Finally, the propagation of the fundamental wave in the increasing $z$ direction is ensured
by also adding the source terms to the magnetic field \cite{devauxOpEx2013}.

In Eq. \ref{eq1}, the electric field components of the fundamental and the SH waves are updated, using the full coupled wave equations describing the SHG process in the LN crystal, as follows:
\begin{eqnarray}\label{eq4}\left\{\begin{array}{ccc}
E_{x,f}|^{n+1}_{jkl} = & \frac{D_{x,f}|^{n+1}_{jkl}-\varepsilon_0\left (d_{31}(E_{x,h}|^{n}_{jkl}E_{y,f}^*|^{n}_{jkl}+E_{y,h}|^{n}_{jkl}E_{x,f}^*|^{n}_{jkl})-d_{22}(E_{x,h}|^{n}_{jkl}E_{z,f}^*|^{n}_{jkl}+E_{z,h}|^{n}_{jkl}E_{x,f}^*|^{n}_{jkl})\right)}{\varepsilon_0\varepsilon_{rx,f}|_{jkl}}\\
E_{y,f}|^{n+1}_{jkl} = & \frac{D_{y,f}|^{n+1}_{jkl}-\varepsilon_0\left (d_{31}(E_{x,h}|^{n}_{jkl}E_{x,f}^*|^{n}_{jkl}+E_{z,h}|^{n}_{jkl}E_{z,f}^*|^{n}_{jkl})+d_{33}E_{y,h}|^{n}_{jkl}E_{y,f}^*|^{n}_{jkl}\right)}{\varepsilon_0\varepsilon_{ry,f}|_{jkl}}\\
E_{z,f}|^{n+1}_{jkl} = & \frac{D_{z,f}|^{n+1}_{jkl}-\varepsilon_0\left (d_{31}(E_{y,h}|^{n}_{jkl}E_{z,f}^*|^{n}_{jkl}+E_{z,h}|^{n}_{jkl}E_{y,f}^*|^{n}_{jkl})-d_{22}(E_{x,h}|^{n}_{jkl}E_{x,f}^*|^{n}_{jkl}-E_{z,h}|^{n}_{jkl}E_{z,f}^*|^{n}_{jkl})\right)}{\varepsilon_0\varepsilon_{rz,f}|_{jkl}}\\
\end{array}\right.
\end{eqnarray}
\begin{eqnarray}\label{eq5}\left\{\begin{array}{ccc}
E_{x,h}|^{n+1}_{jkl} = & \frac{D_{x,h}|^{n+1}_{jkl}-\varepsilon_0\left (2d_{31}E_{x,f}|^{n}_{jkl}E_{y,f}|^{n}_{jkl}-2d_{22}E_{x,f}|^{n}_{jkl}E_{z,f}|^{n}_{jkl}\right )}{\varepsilon_0\varepsilon_{rx,h}|_{jkl}}\\
E_{y,h}|^{n+1}_{jkl} = & \frac{D_{y,h}|^{n+1}_{jkl}-\varepsilon_0\left (d_{31}(E_{x,f}^2|^{n}_{jkl}+E_{z,f}^2|^{n}_{jkl})+d_{33}E_{y,f}^2|^{n}_{jkl}\right ) }{\varepsilon_0\varepsilon_{ry,h}|_{jkl}}\\
E_{z,h}|^{n+1}_{jkl} = & \frac{D_{z,h}|^{n+1}_{jkl}-\varepsilon_0\left (d_{22}(-E_{x,f}^2|^{n}_{jkl}+E_{z,f}^2|^{n}_{jkl})+2d_{31}E_{y,f}|^{n}_{jkl}E_{z,f}|^{n}_{jkl}\right) }{\varepsilon_0\varepsilon_{rz,h}|_{jkl}}\\
\end{array}\right.
\end{eqnarray}
where $d_{22}$, $d_{31}$ and $d_{33}$ are the quadratic nonlinear coefficients of the LN crystal. These coefficients are set to zero outside the LN medium. The periodic poling is modelled by changing periodically the sign of the coefficient $d_{33}$ (Fig.\ref{schemaridgeSHG}) with the period $\Lambda =L_c$, where $L_c$ is the coherence length of the SHG process. $\varepsilon_{rv,w}|_{jkl}$ is the relative permittivity of the sampled media along the $v$ direction, at the spatial sampling point $(j\Delta x,k\Delta y,l\Delta z)$ and at the wavelength $\lambda_w$ ($w\equiv f,h$). This parameter takes in account the birefringence of the LN crystal.

\section{Numerical results}\label{resultats}
\subsection{ SHG in uniformly poled LN ridge-type waveguides}\label{nonpoledsection}
First, we use the 3D-PSTD algorithm for modeling the SHG process in an uniformly poled waveguide with a square section of $10\times 10\,\mu m^2$ . The source emits a 1$mW$ continuous (CW) gaussian beam at the fundamental wavelength $\lambda_f$ which is linearly polarized along the c-axis of the LN crystal ($\psi_f=\frac{\pi}{2}$, $\varphi_{x,f}=0$ and $\varphi_{y,f}=0$). The gaussian beam profile is adjusted in order to optimize the coupling of the fundamental wave in the waveguide. The other parameters of the simulation are listed in table \ref{table1}.
\begin{table}[h]
\caption{Numerical values of the simulation parameters.}\label{table1}
\begin{center}
\begin{tabular}{c c c}
     \hline\hline    
 Wavelengths (in $\mu m$) & $\lambda_f$, $\lambda_h$ & 1.55, 0.775\\
 LN refractive indices & $n_{of}$, $n_{ef}$, $n_{oh}$, $n_{eh}$ & 2.2128, 2.1371, 2.2606, 2.1780\\
 LN nonlinear coefficients (in $pm/V$)& $d_{22}$, $d_{31}$, $d_{33}$& 2.1, -4.64, -41.7\\
 Sampling points & $n_x\times n_y\times n_z$ & $128\times 128\times 256$ \\
 Spatial sampling steps & $\Delta x=\Delta y=\Delta z$& $\frac{\lambda_h}{2.2n_{eh}}$ \\
 Temporal sampling step &   $\Delta t$ & $\frac{\Delta x}{8c}$\\
 Buffer layer refractive index  & $n_v$ & 1.501\\
 Silicon refractive indices & $n_{Sif}$, $n_{Sih}$& 3.50 , 3.72\\
 Ridge length (in $\mu m$)& L & 30\\ 
 \hline\hline  
\end{tabular}\end{center}
\end{table}

\begin{figure}[ht]
   \begin{center}
   \includegraphics[height=8cm]{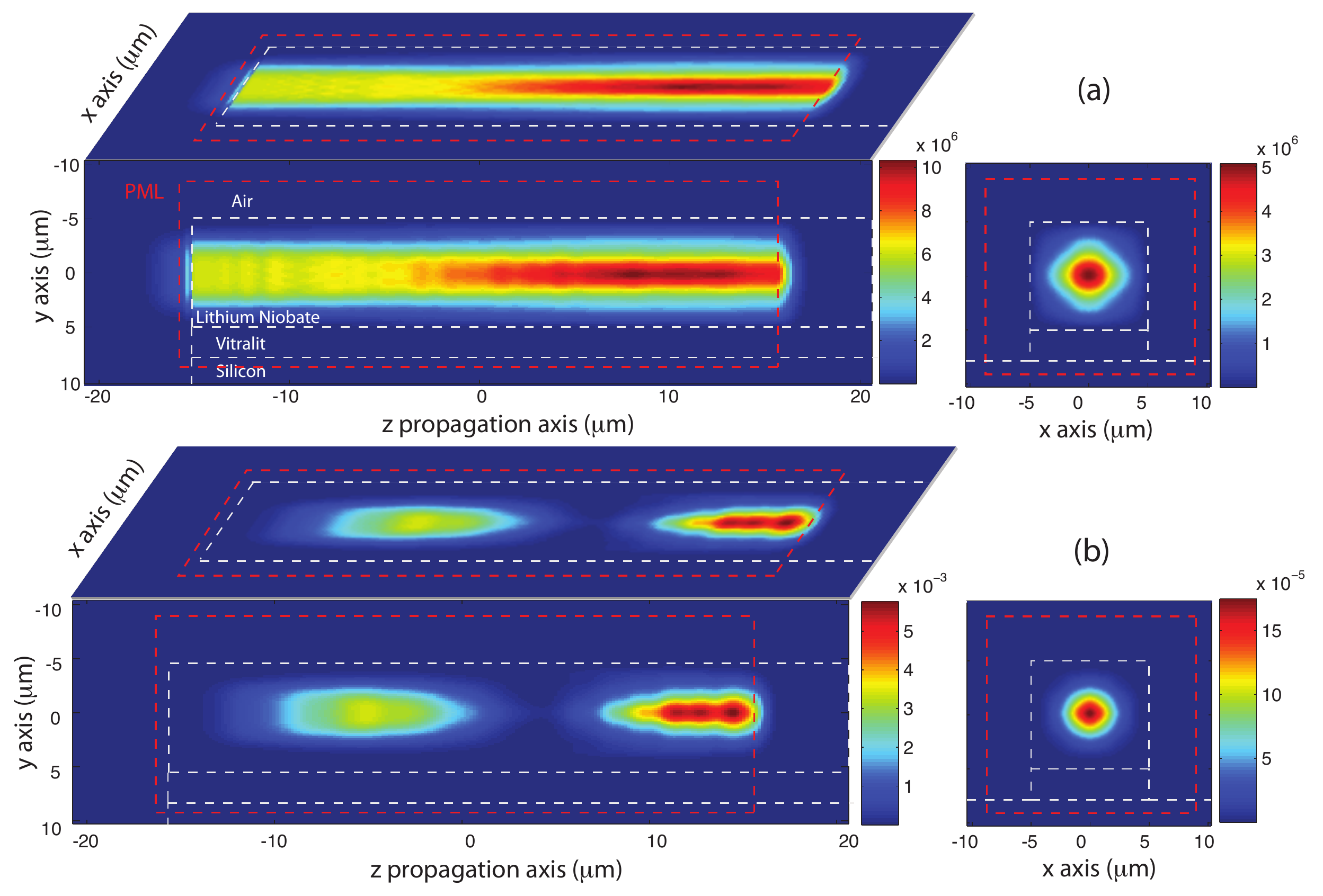}
   \end{center}
   \caption{ \label{simulridge10microns}
Numerical results: (a) propagation of the fundamental wave inside a ridge-type waveguide with a square section of $10 \times 10\,\mu m^2$. (b) SH wave generated in the ridge. Dotted red lines show the position of the the perfectly phase matched layers boundaries of the sampled volume. Dotted white lines denote the interfaces between the different materials.}
\end{figure}

Figure \ref{simulridge10microns} depicts the intensity distribution along the ridge waveguide and at the output for both the fundamental and the SH waves. Dotted red lines give the position of the perfect phase matched layers (PML) at the boundaries of the sampled volume and dotted white lines correspond to the interfaces between the different materials (air, LN, buffer and silicon). Along the $30 \mu m$ propagation length many features can be observed. First, over about $10 \mu m$ length the launched beam is reshaped to form the guided beam mainly constituted of the fundamental mode. The oscillatory behavior of the SH beam intensity is also clearly seen with about two periods observed. At the output of the waveguide a wider beam size is obtained for the fundamental beam compare to the SH beam as expected from the wavelength difference. We would like to emphasize that the decay of the fundamental wave intensity at the end of the waveguide is due to the absorbing layer. Numerical simulations have also been performed for waveguides with $5 \times 5$ and $2.5 \times 2.5\,\mu m^2$ square sections.

\begin{figure}[ht]
   \begin{center}
   \includegraphics[width=8cm]{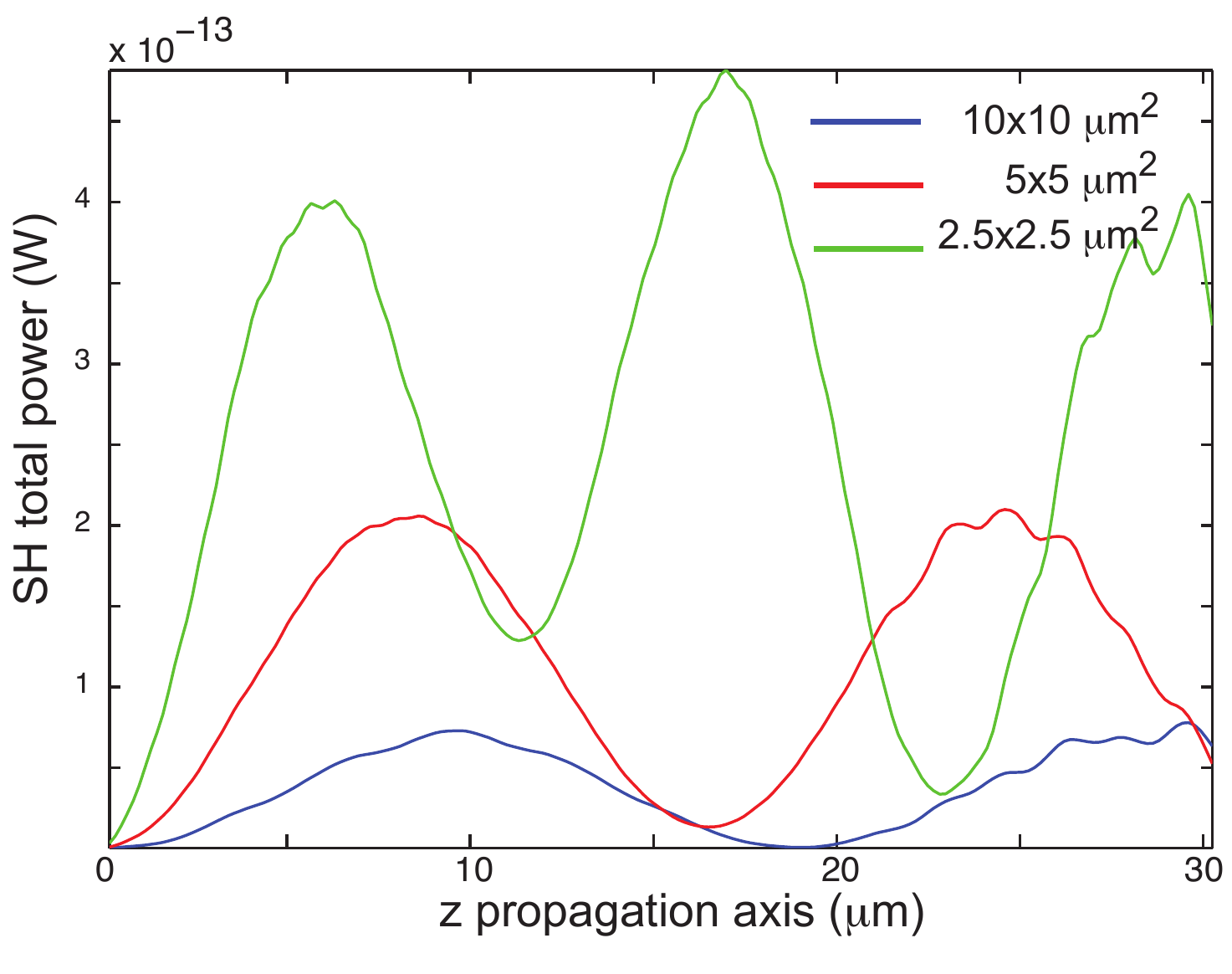}
   \end{center}
   \caption{ \label{plotIzhsanspoling}
For the different modeled guides, total power of the SH generated in the non poled LN ridge-type waveguides as a function of the propagation distance.}
\end{figure}

From the variation of the total power of the SH wave as a function of the propagation distance (Fig. \ref{plotIzhsanspoling}), the coherence length of the SHG process can be estimated with a fairly good accuracy. Note that the oscillatory behavior is less and less perfect as the section of the waveguide is deacresed. Indeed, intensity variation of the SH wave along the propagation axis in the larger waveguide is almost sinusoidal in good agreement with the usual plane wave formalism while this is not the case for smaller waveguides even when size of the input gaussian beam is optimized. These discrepancies are due to the reshaping distance as well as the presence of weakly excited higher order modes. This latter point is confirmed by numerical simulations that show stronger intensity fluctuations of fundamental guided waves along the propagation axis in smaller waveguides. This suggests that injected light is coupled with higher order guided modes which implies that different harmonic frequency waves with different coherence lengths are generated simultaneously.   

Coherence lengths estimated with the PSTD simulations can be compared with the coherence lengths calculated with the usual definition:
\begin{equation}\label{eq6}
L_c=\frac{\lambda_f}{4(n_{h,eff}-n_{f,eff})}
\end{equation}
where $n_{h,eff}$ and $n_{f,eff}$ are the effective indices of the fundamental eigenmodes at the wavelengths $\lambda_h$ and $\lambda_f$, respectively. For the modeled waveguides, table \ref{table2} gives the coherence lengths deduced from PSTD simulations and the coherence lengths calculated with Eq. \ref{eq6} where effective indices are calculated with a commercial software. Coherence lengths obtained from both methods are in good agreement.     
\begin{table}[h]
\caption{\label{table2}Comparison between the coherence lengths deduced from the PSTD simulations and calculated  with Eq. \ref{eq6}.}
\begin{center}
\begin{tabular}{c c c c}
     \hline\hline    
 Square section of Ridge  (in $\mu m^2$) & $10 \times 10$  & $5 \times 5$ &$2.5 \times 2.5$\\
 Coherence length with PSTD ($\mu m$)& 8.7 & 7.7 & 5.4\\
 LN Effective index $n_{f,eff}$& 2.1340 & 2.1265 & 2.0975\\
 LN Effective index $n_{h,eff}$& 2.1773 & 2.1753 & 2.1677 \\
 Coherence length with Eq. \ref{eq6} ($\mu m$)& 8.9 & 8.2 & 6.0\\
 \hline\hline  
\end{tabular}\end{center}
\end{table}
\subsection{SHG in periodically poled LN ridge-type waveguides}\label{poledsection}
\begin{figure}[t]
   \begin{center}
   \includegraphics[width=15cm]{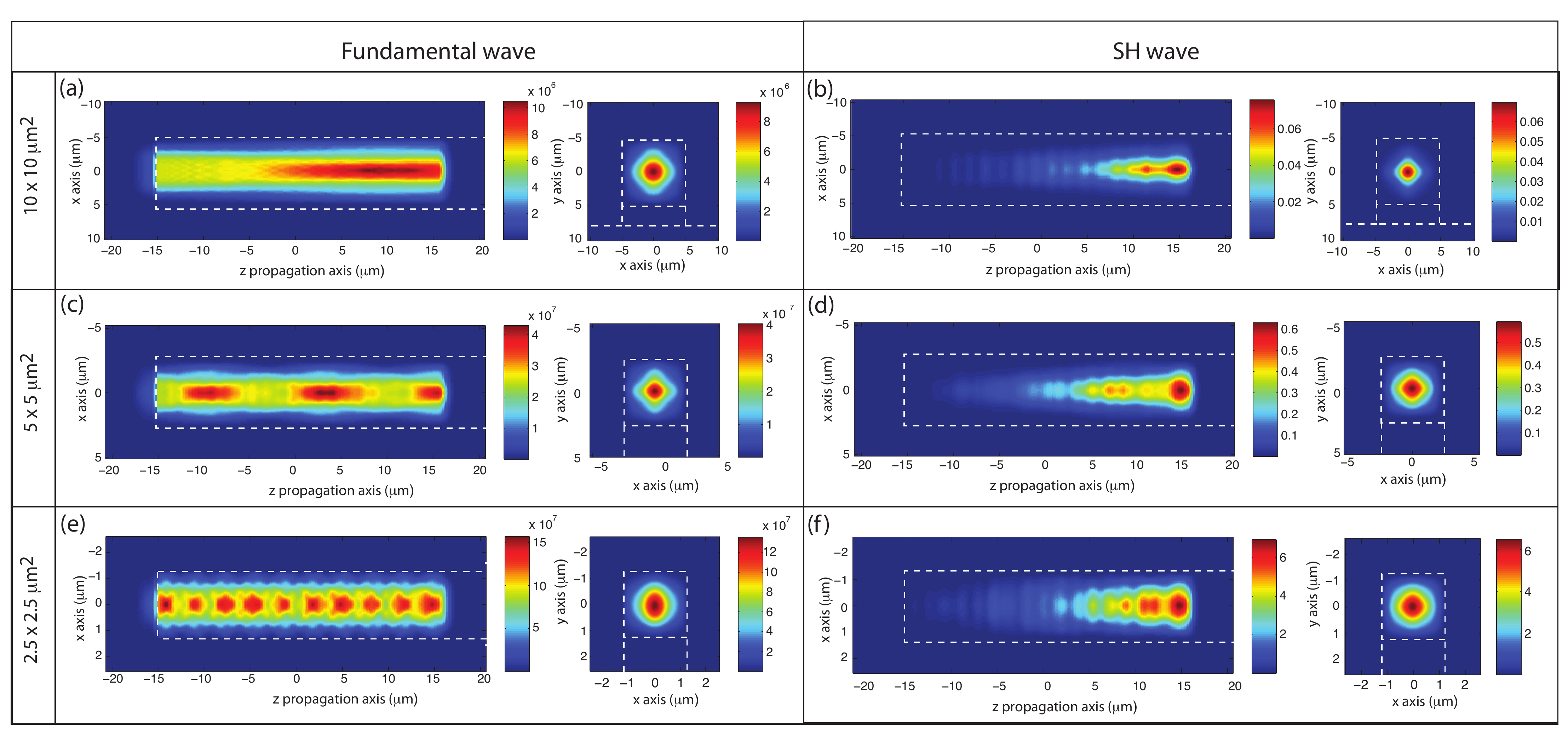}
   \end{center}
   \caption{ \label{figridgeavecpoling}
Numerical results: (a,c,e) propagation of the fundamental wave through the periodically polled ridge-type waveguides with different square section. (b,d,f) SH wave generated in the waveguides.  Dotted white lines denote the interfaces between the different materials.}
\end{figure}
Next, the coherence lengths deduced from the above PSTD simulations (table \ref{table2}) are used to design ridge-type waveguides where domains are now periodically poled with a period twice the coherence length. A series of simulations are then performed for the three considered structures. Figure \ref{figridgeavecpoling} shows the intensity distribution, in the $xz$ and $xy$ planes, of the fundamental and SH waves. For all square sections the transverse spatial sampling steps ($\Delta x$ and $\Delta y$) is adjusted such as the total number of sampling points in the transverse plane is a constant. We observe that waveguides with smaller cross sections (Fig. \ref{figridgeavecpoling}c and \ref{figridgeavecpoling}e) show intensity fluctuations associated with shorter periods along propagation. Such fluctuations denote the multimode character of the guided waves eventhough the size of the launched gaussian beam is optimized to excite mainly the fundamental mode. As the waveguide section is reduced the mode beating is associated with shorter period which is in agreement with effective indices of modes that are more and more dissimilar. This trend is in accordance with modal properties of waveguides. As a consequence, SH waves also suffer from noise due to both the multimode and the phase matching conditions.
\begin{figure}[ht]
   \begin{center}
   \includegraphics[width=8cm]{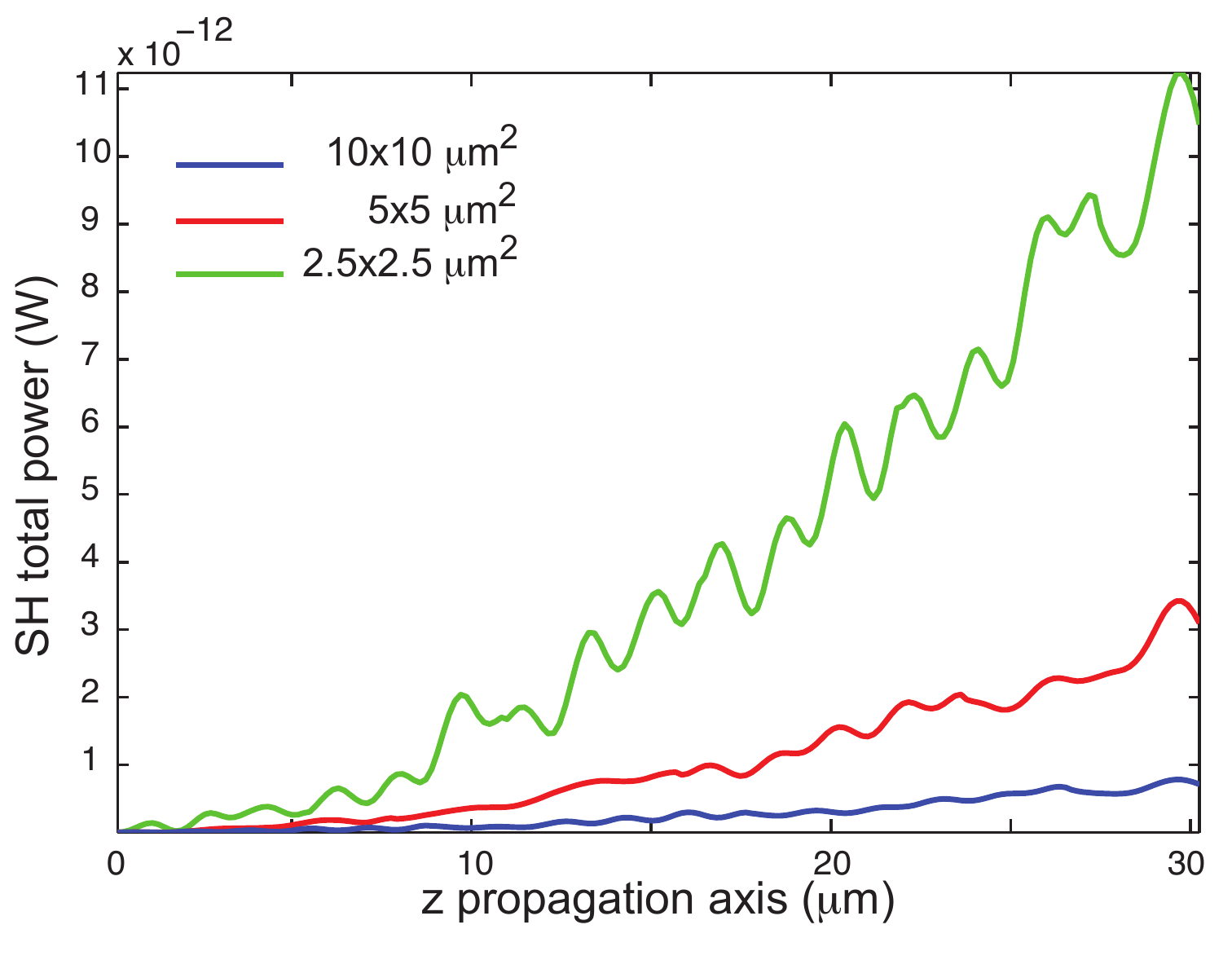}
   \end{center}
   \caption{ \label{plotIzavecpoling} Total power of the SH generated waves along propagation distance in the periodically poled LN ridge-type waveguides of three different cross sections.}
\end{figure}
Fig. \ref{plotIzavecpoling} shows the evolution of the total power of the SH wave as a function of the propagation distance. Since the total power of the launched source is a constant, the power density of the guided beam increases when square section of the guide decreases. Consequently, the nonlinear phenomena is more efficient and the total power of the harmonic wave increases. According to the basic theory, this gain in generated power should increases like the inverse of the section of the waveguide. This trend is present in the depicted results but a slightly lower gain is obtained that can partialy be attributed to the presence of the beam reshaping length.   

\begin{figure}[ht]
   \begin{center}
   \includegraphics[width=8cm]{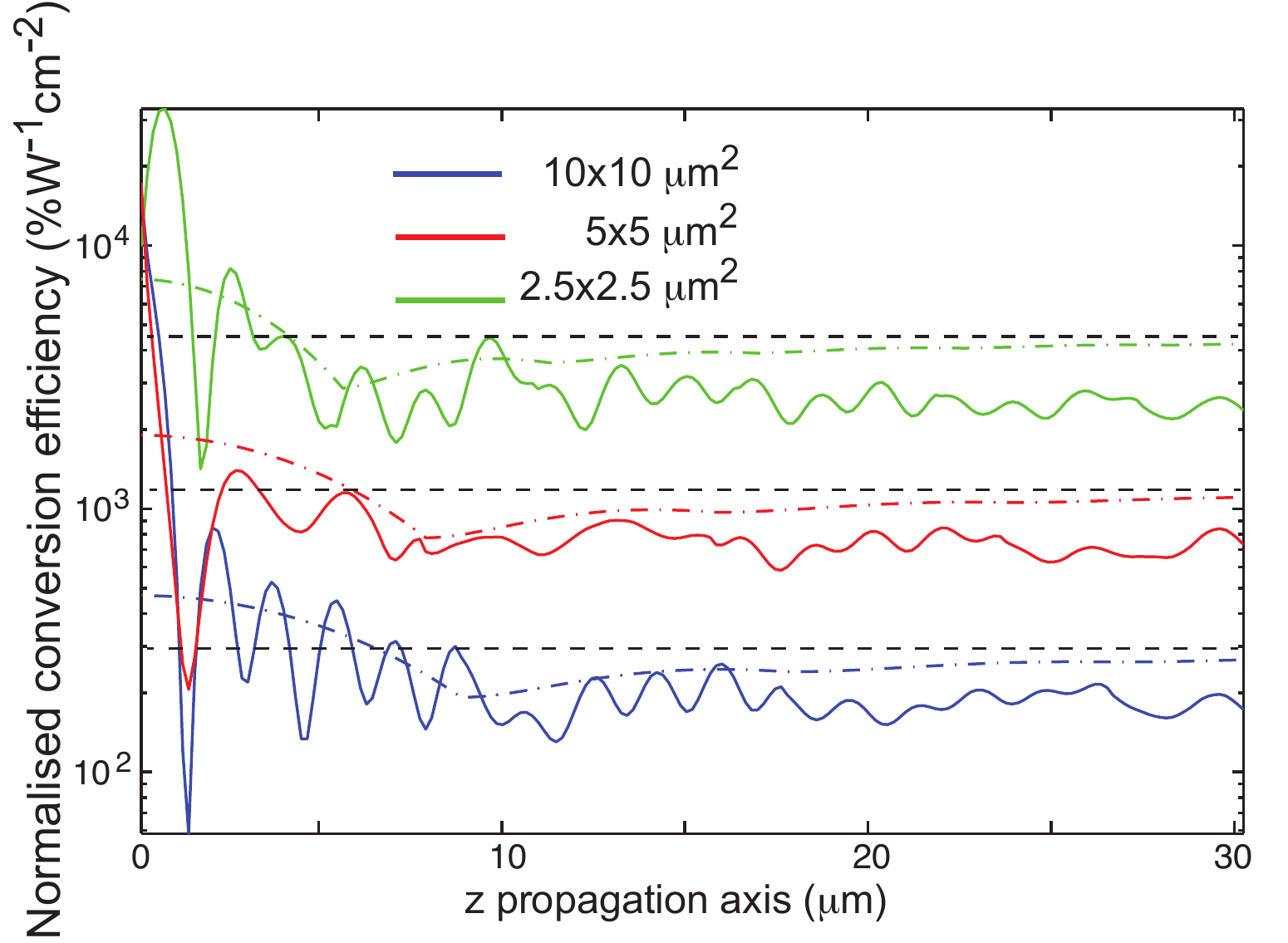}
   \end{center}
   \caption{ \label{plotrendementnormavecpoling} Normalized SHG efficiencies in the periodically poled LN ridge-type waveguides. Color dotted curves show the corresponding theoretical normalized efficiencies calculated with Eq. \ref{eq8}. Black dotted lines symbolise for each ridge the efficiency level $\frac{2\eta_{max}}{\pi}$.}
\end{figure}

In order to estimate the efficiency of the non linear process we plot the normalized efficiencies (in $ \% W^{-1}cm^{-2}$) of the SHG process (Fig. \ref{plotrendementnormavecpoling}) as a function of the propagation length which is defined as: 
  \begin{equation}\label{eq7}
  \eta=\frac{P_h}{P^2_fz^2},
  \end{equation} 
where $P_h$ and $P_f$ are the total power of the SH and fundamental waves inside the waveguides, respectively and $z$ is the propagation distance. To compare with, the dotted curves correspond to the theoretical efficiencies defined for the first domain as :
\begin{equation}\label{eq8}
\eta_{th}(z)=\frac{\Sigma_h}{\Sigma_f^2}\frac{2w_f^2d^2_{33}}{n_{h,eff}n^2_{f,eff}c^3\epsilon_0}sinc^2\left(\frac{\Delta k z}{2}\right)=\eta_{max}sinc^2\left(\frac{\Delta k z}{2}\right),\end{equation} 
where $\Sigma_h$, $\Sigma_f$ are the cross section areas of the guided beam and $\Delta k=\frac{4\pi}{\lambda_f} \left(n_{h,eff}-n_{f,eff}\right)$ is the phase mismatch between the fundamental and the SH waves and $\eta_{max}$ is the efficiency for perfect phase matching. Because of the periodic poling of the waveguide, efficiency along the $p^{th}$ domain (with $p>1$) is calculated by incrementing the efficiency at the end of the domain $p-1$ as follows:  
\begin{equation}\label{eq9}
    \eta_{th}\left(z\right)=\eta_{th}\left((p-1)L_c\right)+\eta_{max}\left(\frac{\sin\left(\frac{\Delta k z}{2}-(p-1)\frac{\pi}{2}\right)}{\frac{\Delta k z}{2}}\right)^2,
    \end{equation} 
where an additional phase shift of $-\frac{\pi}{2}$ is applied to the sinus function from one domain to the next.
These curves show that the normalised efficiency, deduced both from the PSTD simulations and from Eq. \ref{eq9}, is high at the beginning of the waveguide and decreases along propagation in the first domain. Then, it rapidly stabilises to an almost constant value. This behavior is inherent to the quasi-phase matching configuration. The reached level is well approximated the efficiency $\eta_{max}$ reduced by a factor of $\frac{2}{\pi}$ as shown in \cite{rustagi_1982}. This level is represented in Fig. \ref{plotrendementnormavecpoling} for each ridge by the black dotted lines. Note that the conversion efficiency calculated from Eq. \ref{eq9} is slightly higher than the one deduced from the PSTD results since the latter method takes into account the multimode character of the waveguide. More importantly, these curves show that very strong conversion efficiency can be obtained when the size of the waveguide decreases.  

\subsection{Modal analysis of the guided waves}\label{modalanalysis}
\begin{figure}[ht]
   \begin{center}
   \includegraphics[width=15cm]{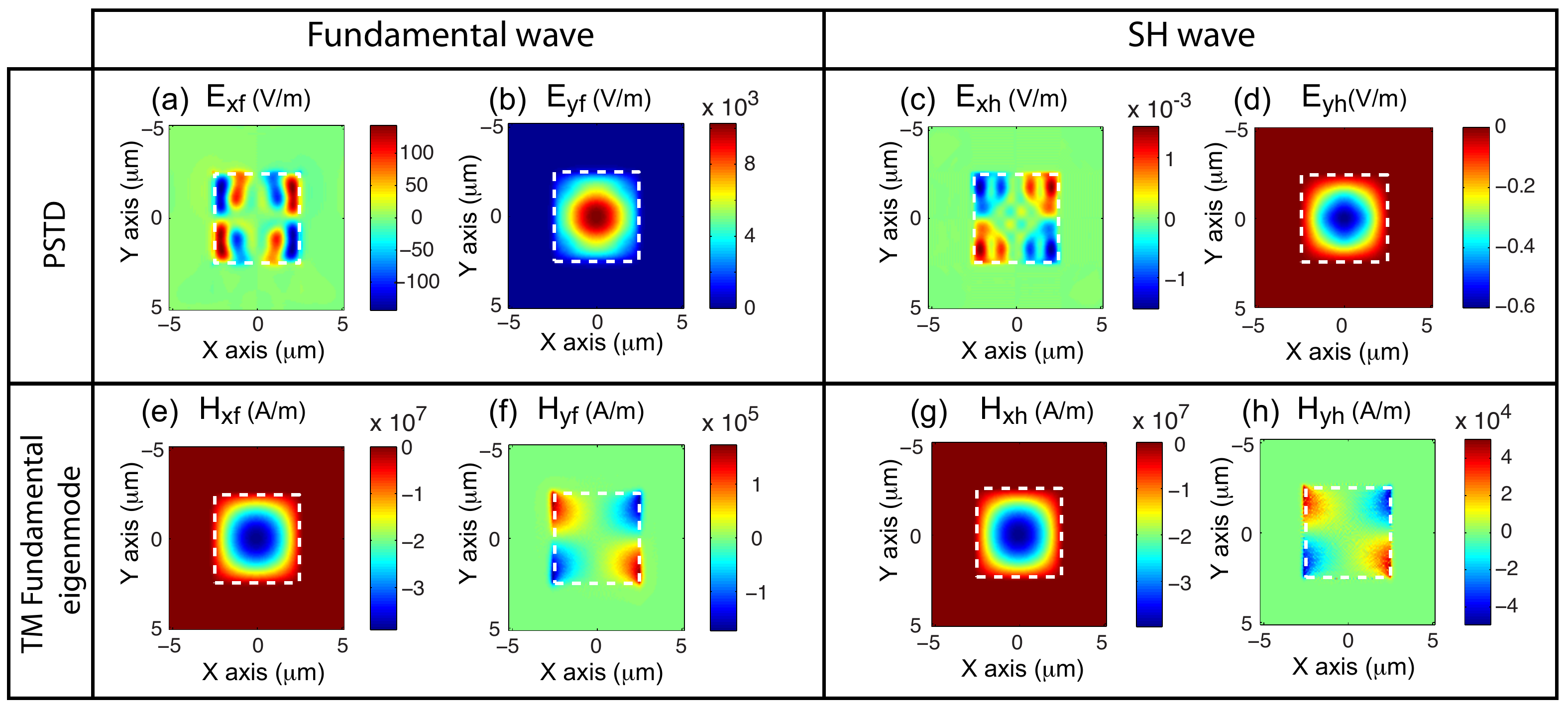}
   \end{center}
   \caption{ \label{PSTDvsCOMSOL}
Field components of the fundamental and harmonic waves for the $5\times5\,\mu m^2$ ridge waveguide : (a-d) electric fields of the output guided waves from the PSTD algorithm and (e-h) magnetic fields of the TM fundamental eigenmodes given by the commercial software. Dotted white squares show the ridge contours.}
\end{figure}

In this last section we study the influence of the multimode property of the ridge waveguides on the SHG process. We first exploit the numerical results given by the PSTD algorithm in periodically poled waveguides in order to evaluate the dominant excited eigenmode. To this purpose, a commercial software is first used to calculate the components of the transverse eigenmodes for each ridge waveguide at both fundamental and harmonic wavelengths. Then, for a given eigenmode $m$ at the wavelength $\lambda_w$ the overlap integral with the output beam obtained with the PSTD method is calculated using :
\begin{equation}\label{eq10}
    \varGamma_w(m)=\frac{\rvert\iint \left(E^{PSTD}_{x,w}H^*_{y,wm}-E^{PSTD}_{y,w}H^*_{x,wm}\right)dxdy\rvert^2}{\iint\rvert E^{PSTD}_w\rvert^2 dxdy\iint\rvert H_{wm}\rvert^2 dxdy},
    \end{equation}
where $(E^{PSTD}_{x,w},E^{PSTD}_{y,w})$ are the electric field components given by the PSTD algorithm and $(H_{x,wm},H_{y,wm})$ are the magnetic field components of the eigenmode $m$ calculated with the commercial software. As an example, Fig. \ref{PSTDvsCOMSOL} shows the transverse components of the electric fields for the fundamental and the harmonic waves at the output of the $5\times5\,\mu m^2$ waveguide along with and the magnetic fields components of the $TM_{00}$ eigenmodes.  
The obtained values of the overlap integral between the guided beam and the first mode ($TM_{00}$) at the fundamental wavelength are superior to 98$\%$ whatever the geometry of the waveguides. It thus shows that the guided mode is mainly composed of the first mode which is a favorable situation to reach high conversion efficiency. The residue of higher order modes still gives some intensity fluctuation due to mode beating which is noticeable for small section waveguides for which beating length is shorter ( see Fig. \ref{figridgeavecpoling}). Additional overlap integral calculations between the fundamental mode of the SH waves and the guided beam at the fundamental wavelength obtained from PSTD show that the overlap decay as the section of the waveguide is getting larger. For instance, the overlap for the $10\times10\,\mu m^2$ ridge is 90$\%$ while it reaches 99$\%$ for the $2.5\times2.5\,\mu m^2$ one. Such an almost perfect matching comes from the strong similarity between the fundamental modes at both wavelengths as observed in Fig. \ref{PSTDvsCOMSOL}. This feature is an advantage of high index contrast ridge type waveguides. We can assert that the strong second harmonic generation efficiency reachable in narrow waveguides is not only due to a tight confinement but it also benefits due to a better mode matching between fundamental modes at both wavelengths. 
\begin{figure}[h]
   \begin{center}
   \includegraphics[width=8cm]{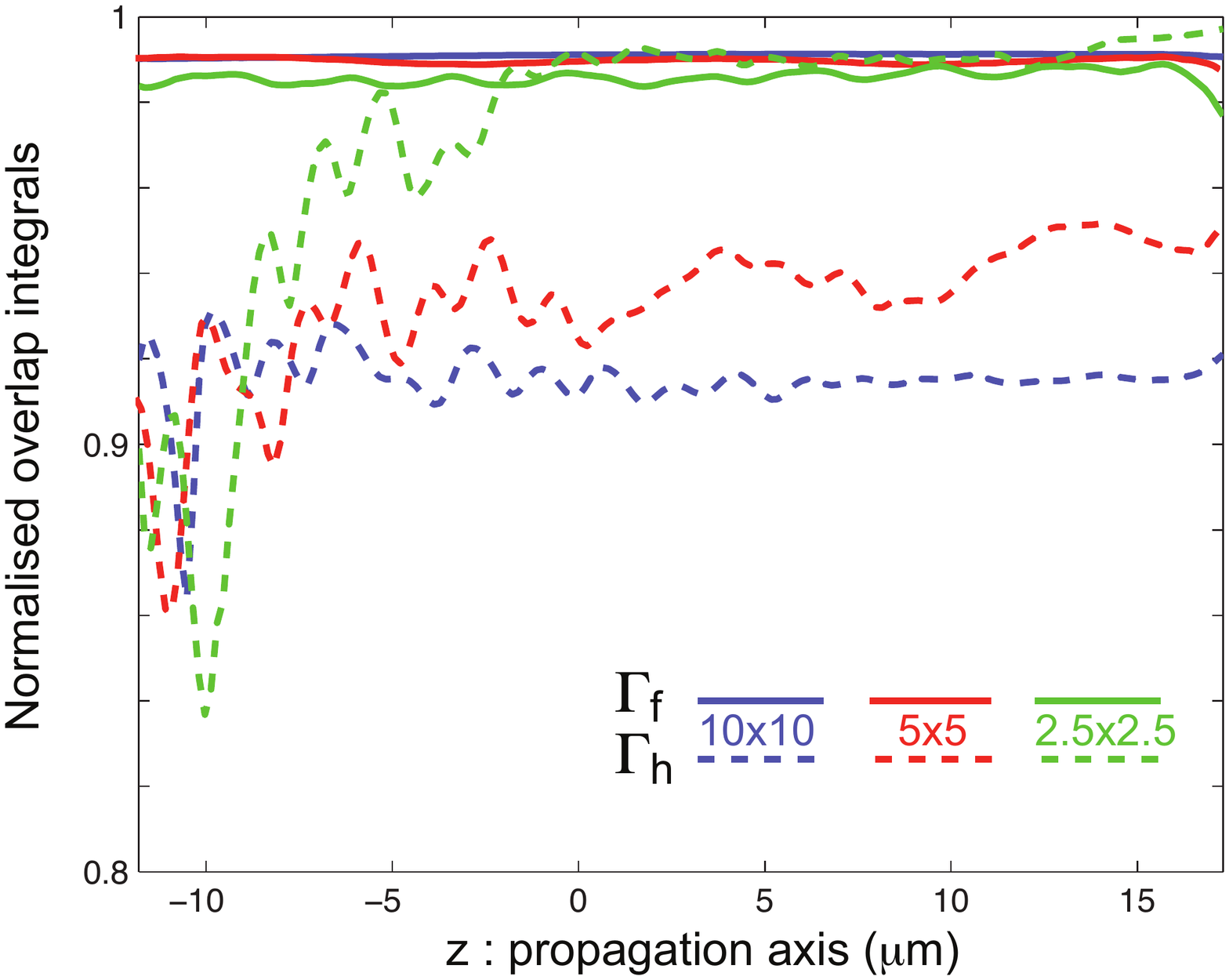}
   \end{center}
   \caption{ \label{overlap}
Variation of the normalised overlap integrals along the propagation axis. Overlap integrals are calculated between the guided light from PSTD and the first eigenmodes at the fundamental (solid curves) and harmonic (dotted curves) wavelengths for the $10\times10$, $5\times5$ and $2.5\times2.5\,\mu m^2$ waveguides.}
\end{figure}

In order to complement these results, we also calculate the overlap integrals along the propagation axis. Figure \ref{overlap} shows the variation of the overlap integrals, between the guided waves and the $TM_{00}$ eigenmodes of the waveguides, along the propagation axis. We note that the launched Gaussian beam shape is optimised as witnessed by an overlap close to 100$\%$ with the fundamental wave starting from the entrance of the waveguides. To the contrary, coupling with the $TM_{00}$ modes of the SH waves exhibits strong fluctuations at the early stage of the propagation and increases quickly up to the optimal values. This effcet is  especially visible in small section waveguides. We attribute the initial stage to the distance necessary to the reshaping and redistribution of the generated SH light among the different eigenmodes.          

\section{Conclusion}
For the first time to our knowledge, a 3D-PSTD algorithm has been implemented to solve the second harmonic generation process in lithium niobate ridge-type waveguides. The model has first been validated with uniformly poled waveguides and further used to characterise and better comprehend the physics in periodically poled ridges. The model is able to determine most characteristics of the SHG process such as optimum poling period, conversion efficiency along with accurate guided beams profile evolution. More specifically, it shows that very high conversion efficiency can be reached in high-index contrast ridge type waveguide even though they are not single mode. These performances are obtained with injection of a gaussian beam that is shaped to favour excitation of the fundamental mode of the guiding structure. For instance, conversion efficiency as high as 2000 $ \% W^{-1}$ is expected in a 1$cm$ long ridge waveguide of $2.5 \times 2.5\,\mu m^2$ square section.

\end{document}